\documentclass[aps,reprint,superscriptaddress]{revtex4-1}
\usepackage{graphicx,amssymb}

\begin{document}
\title{Unidirectional spin-wave channeling along magnetic domain walls of Bloch type}
\author{Y. Henry}
\email{yves.henry@ipcms.unistra.fr}
\affiliation{Universit\'e de Strasbourg, CNRS, Institut de Physique et Chimie des Mat\'eriaux de Strasbourg, UMR 7504, 67034 Strasbourg, France}
\author{D. Stoeffler}
\affiliation{Universit\'e de Strasbourg, CNRS, Institut de Physique et Chimie des Mat\'eriaux de Strasbourg, UMR 7504, 67034 Strasbourg, France}
\author{J.-V. Kim}
\affiliation{Centre de Nanosciences et de Nanotechnologies, CNRS, Univ. Paris-Sud, Universit\'e Paris-Saclay, 91405 Orsay, France}
\author{M. Bailleul}
\affiliation{Universit\'e de Strasbourg, CNRS, Institut de Physique et Chimie des Mat\'eriaux de Strasbourg, UMR 7504, 67034 Strasbourg, France}
\date{\today}

\begin{abstract}
From the pioneering work of Winter [Phys. Rev. \textbf{124}, 452 (1961)], a magnetic domain wall of Bloch type is known to host a special wall-bound spin-wave mode, which corresponds to spin-waves being channeled along the magnetic texture. Using micromagnetic simulations, we investigate spin-waves travelling inside Bloch walls formed in thin magnetic media with perpendicular-to-plane magnetic anisotropy and we show that their propagation is actually strongly nonreciprocal, as a result of dynamic dipolar interactions. We investigate spin-wave non-reciprocity effects in single Bloch walls, which allows us to clearly pinpoint their origin, as well as in arrays of parallel walls in stripe domain configurations. For such arrays, a complex domain-wall-bound spin-wave band structure develops, some aspects of which can be understood qualitatively from the single-wall picture by considering that a wall array consists of a sequence of up/down and down/up walls with opposite non-reciprocities. Circumstances are identified in which the non-reciprocity is so extreme that spin-wave propagation inside individual walls becomes unidirectional.
\end{abstract}

\maketitle

Nonreciprocal spin-wave propagation is well-known to occur in transversally magnetized thin films. In the so-called Damon-Eshbach (DE) configuration, indeed, spin-waves have a surface character and their localization changes from one surface to the other under inversion of the direction of propagation\,\cite{DE61}. As a result, counter-propagating spin-waves experience inequivalent effective media and thus behave differently as soon as the film possesses vertically asymmetric magnetic properties. In particular, counter-propagating spin-waves with a given frequency (resp. wave vector) have different wave vectors (resp. frequencies)\,\cite{H90,K13,GHHK16}. Channeling of spin-waves along magnetic domain walls, on the other hand, was first predicted in the early 1960's [Ref.~\onlinecite{W61}]. Recently, it has been reconsidered, both theoretically\,\cite{GBSA15,LYWX15,XZ16} and experimentally\,\cite{WKSH16,APSS17,SSGK18}, after physicists realized that it could provide an efficient way to guide spin-waves along curved and/or reconfigurable paths in magnonic circuits\,\cite{GBSA15} as well as for building reprogrammable logic hardware architectures\,\cite{LYWX15}. So far, nonreciprocal channeling of spin-waves has essentially been observed along chiral N\'eel walls\,\cite{GBSA15,LYWX15,XZ16}, as well as edge spin textures\,\cite{GBVK14}, all of them being stabilized by an interfacial Dzyaloshinskii-Moriya (iDM) interaction in ultrathin out-of-plane magnetized media. Following the seminal work by Winter\,\cite{W61}, Bloch walls have been considered as hosting fully reciprocal domain-wall channeled spin-wave (DWCSW) modes\,\cite{GBSA15}. However, recent theoretical works by Makhfudz and collaborators\,\cite{MKT12} and Zhang and Tchernyshyov\,\cite{ZT18}, based on a general Lagrangian approach, suggest that such domain walls actually exhibit very unusual dynamics and behave like nonreciprocal strings, where transverse waves propagating in opposite directions have different velocities.

In the present paper, we study nonreciprocal propagation of spin-waves inside Bloch walls in details using micromagnetic simulations performed both with the MuMax3 program\,\cite{VLDH14,MuMax3} and with an in-house developed numerical approach based on the dynamic matrix method, which we will refer to as SWIIM, and which allows performing a spin-wave normal mode analysis\,\cite{HGB16}. First, for the purpose of comparison with a multi wall situation and for describing the effect with the usual spin-wave vocabulary, we revisit the case of a single wall. We show that, because the magnetic equilibrium in the presence of a Bloch wall bears some resemblance with the DE configuration and is not symmetric about its midplane, the DWCSW mode bound to a Bloch wall is indeed intrinsically and sometimes very strongly nonreciprocal. Here, spin-wave non-reciprocity is not produced by the chiral iDM interaction but by dynamic dipolar interactions, just as for asymmetric films in the surface wave configuration. Next, we go one step further and consider DWCSW modes in arrays of parallel Bloch walls, in stripe domain configurations. We verify that there exist as many of these DWCSW modes as there are domain walls and we show that the complex magnonic band structure formed can be partly understood from the single wall picture by considering that the wall array consists of a sequence of up/down and down/up walls with opposite non-reciprocities. At low wave vectors, dipolar coupling between neighboring walls combined with finite-size effects lead to the formation of avoided crossings and a hierarchy of regularly-spaced spin-wave branches. At large wave vectors, the domain walls become dipolar decoupled and the DWCSW bands form two distinct groups, with all the modes in a group being almost degenerate in frequency and corresponding to spin-wave propagating exclusively inside domain walls of a particular type, i.e., in one in every two walls. We also discuss the response of arrays of dipolar decoupled Bloch walls to localized magnetic field excitations of various frequencies and we show that there exist conditions for which the propagation of domain-wall channeled spin-waves is purely unidirectional, meaning that domain walls act as spin-wave diodes.

\begin{figure}[t]
\includegraphics[width=8.5cm,trim=30 115 30 117,clip]{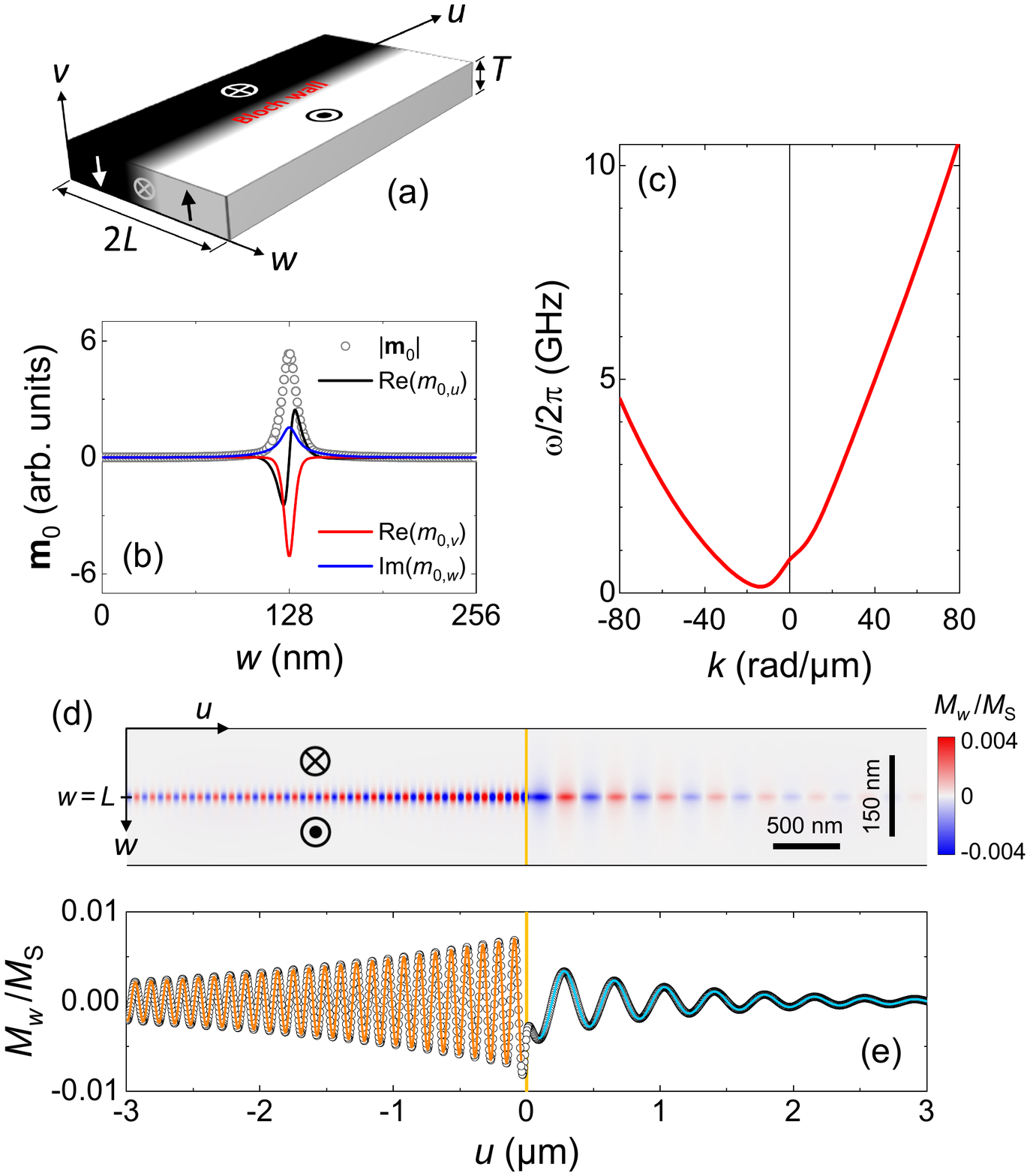}
\caption{(a) Schematic of the magnetic medium containing two stripe domains separated by a Bloch wall ($W\!=\!2L\!=\!256$~nm, $T\!=\!15$~nm). (b) Width profile of the variable magnetization in the lowest frequency spin-wave eigenmode, which propagates as a plane wave $\mathbf{m}(u,w,t) = \mathbf{m}_{0}(w)\, e^{i(\omega t-ku)}$ with $k\!=\!+50$~rad/$\mu$m, along the medium ($u$). (c) Dispersion relation of the DWCSW mode. (d) Map of the transverse component of the magnetization, $M_w$, as generated by an excitation having the form of a 5~mT, vertically oriented, sinusoidal magnetic field with frequency $\omega/2\pi=2$~GHz, localized in $u\!=\!0$ (yellow line). Note the different scales along the $u$ and $w$ directions. (e) Line profile of $M_w$ at the equilibrium position of the domain wall center ($w\!=\!L$). The data (symbols) can be fitted (lines) to the expression $M_w(u) = M_{0,w}\, \sin[k(u\!-\!u_0)]\, e^{-(u-u_0)/L_{\text{att}}}$, with $k=16.7$~rad/$\mu$m and $L_{\text{att}}=1.00~\mu$m (resp. $k\!=\!-53.1$~rad/$\mu$m, $L_{\text{att}}\!=\!-2.42~\mu$m) for channeled spin-waves propagating to the right (resp. to the left).} \label{Fig_One_Wall}
\end{figure}

We consider thin magnetic strips with extremely large length to width aspect ratio, made of an hypothetical material with gyromagnetic ratio $|\gamma/2\pi|=28.02$~GHz/T, damping factor $\alpha=0.01$, saturation magnetization $M_{\text{S}}=1$~MA/m, exchange constant $A=15$~pJ/m, and a strong perpendicular-to-plane uniaxial magnetic anisotropy of constant $K=1$~MJ/m$^3$ ($Q=2K/\mu_0M_{\text{S}}^2=1.6$). As a result of the competition between anisotropy and demagnetizing energies, such media with $Q$ factor larger than unity naturally host stripe domains, that is, elongated domains with magnetization pointing perpendicular to the plane, alternatively up and down, separated by domain walls, all of which are parallel to the (long) edges of the medium. For a thickness $T=15$~nm, as considered throughout this entire work, the equilibrium width of the domains in a fully extended film of the same material is $L=128$~nm and the internal structure of the walls is very close to the Bloch geometry, where magnetic moments lie and rotate in a plane parallel to the wall and magnetization orientation does not vary across the thickness. Therefore, the widths of the media we discuss below are multiples of $L$, $W\!=\!(N+1)L$ where N is the number of walls, and a single magnetic cell is used for discretization in the thickness direction. SWIIM simulations further assume infinitely long magnetic media discretized transversally with 0.2-1~nm wide finite-difference cells. MuMax3 simulations, on the other hand, are performed for 65536~nm long strips discretized with 1~nm wide, 4~nm long cells. Periodic boundary conditions are applied along the direction of the domain wall(s) and spin-wave excitation is induced by out-of-plane continuous-wave magnetic fields applied in a 8~nm long region located half way along the strips and spanning their entire width. Provided that the MuMax3 simulations are carried out in the linear regime, the two numerical methods yield perfectly consistent results and, in particular, the very same frequency versus wave vector dependencies. SWIIM is preferentially used to obtain the complete set of normal modes and their spatial profiles, whereas MuMax3 is employed to generate magnetization maps under specific excitation conditions.

We first investigate the case of a single Bloch wall in a strip with $W\!=\!2L\!=\!256$~nm, containing two oppositely magnetized domains [Fig.~\ref{Fig_One_Wall}(a)], with magnetization pointing down in the leftmost domain ($w<L$) and up in the rightmost one ($w>L$). In zero applied magnetic field, the wall naturally sits at the center of the strip. In agreement with previous works\,\cite{W61,GBSA15}, a normal mode analysis reveals that a special spin-wave mode is bound to the Bloch wall. The latter lies at much lower frequency than bulk-like modes and corresponds to one-dimensional propagation of spin-waves along the wall [Fig.~\ref{Fig_One_Wall}(b)]. The dispersion relation of such a DWCSW mode is markedly asymmetric with respect to $k\!=\!0$ [Fig.~\ref{Fig_One_Wall}(c)]. This means that leftward and rightward propagating channeled spin-waves with a given frequency can actually have very different wavelengths [Fig.~\ref{Fig_One_Wall}(d,e)]. There even exists a range of frequencies ($f<0.8$~GHz) where counter-propagating waves have radically different nature: those with positive group velocity $v_g=\frac{\partial w}{\partial k}$, which propagate to the right, have negative phase velocity $v_p=\frac{w}{k}$ and are therefore backward waves, whereas those propagating to the left ($v_g<0$) are forward waves [$\text{sgn}(v_g)\!=\text{sgn}(v_p)$].

\begin{figure}[t]
\includegraphics[width=8.5cm,trim=26 207 28 170,clip]{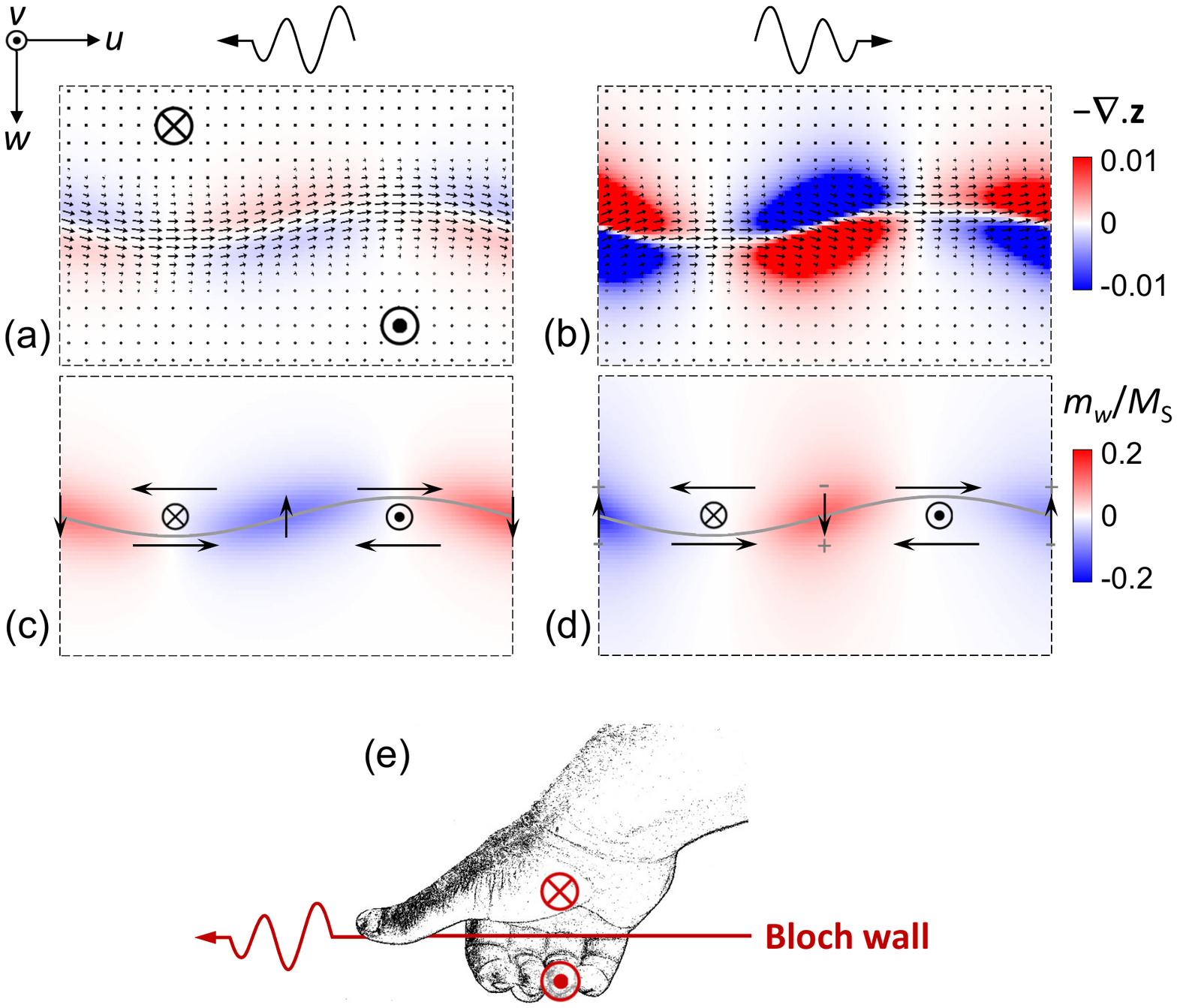}
\caption{(a,b) Close views of the magnetic configurations (arrows) and normalized magnetic charge distributions $\rho\!=\!-\nabla\!\cdot\!\mathbf{z}$ (color map), where $\mathbf{z}\!=\!\mathbf{M}/M_{\text{S}}$, for (a) leftward propagating ($\omega/2\pi\!=\!0.61$~GHz) and (b) rightward propagating ($\omega/2\pi\!=\!3.66$~GHz) channeled spin-waves with wave vector $|k|=30$~rad/$\mu$m. Different (large) strengths of the localized excitation field have been used in order to obtain (clearly visible) domain wall flexures of comparable amplitudes in the two cases: (a) 100~mT, (b) 300~mT. (c,d) Map of the transverse component ($m_w$) of the variable magnetization $\mathbf{m}$ in the magnetic configurations shown in (a) and (b). The arrows indicate schematically the location of the extrema of the three components of $\mathbf{m}$ in the DWCSW mode. Note that, for the sake of clarity, the vertical distance between the extrema of $m_u$ has been exaggerated, as a comparison with Fig.~\ref{Fig_One_Wall}(b) reveals. In all panels (a-d), the area shown has a width of 64~nm and a length equal to the spin-wave wavelength $2\pi/|k|=209$~nm. (e) Illustration of the right hand thumb rule for determining in which direction the channeled spin-waves have the lowest frequency.} \label{Fig_Origin_of_nonreciprocity}
\end{figure}

The origin of these non-reciprocity effects, which become more and more pronounced with increasing strip thickness\,\cite{HGB16}, can be safely ascribed to dynamic dipolar interactions. Indeed, the two conditions necessary for dipole induced frequency non-reciprocity to occur\,\cite{C87} are met: i) First, quite similar to the Damon-Eshbach configuration, the system considered here is (partially) magnetized along a direction $v$, which is at right angle to the direction of spin-wave propagation $u$\,\cite{Rem1}; ii) Second, the system is magnetically inhomogeneous in the direction perpendicular to both $u$ and $v$, that is $w$, and it lacks reflectional symmetry about the midplane normal to this direction [$uv$ plane such that $w=L$, see Fig.~\ref{Fig_One_Wall}(a)]. Micromagnetically speaking, counter-propagating channeled spin-waves with the same wave vector $|k|$ are characterized by nonequivalent variable magnetization maps\,\cite{HGB16} and very different spatial distributions of dynamically created volume magnetic charges [Fig.~\ref{Fig_Origin_of_nonreciprocity}(a,b)]: the largest those volume charges, the highest the frequency of the DWCSW mode. The reason for this difference in behavior can be traced back to the fact that these waves have phase velocities with opposite signs. To illustrate this point, let us assume that, at a given time, the out-of-plane component of the variable magnetization $\mathbf{m}$, $m_v$, is the same for the two directions of propagation so that the Bloch wall exhibits identical flexures in both cases [Fig.~\ref{Fig_Origin_of_nonreciprocity}(a,b)] and surface magnetic charges are the same too. In such a situation, a change of sign of $v_p$ is inevitably accompanied by a change of sign of the transverse component of $\mathbf{m}$, $m_w$, which oscillates in quadrature with $m_v$ [Fig.~\ref{Fig_Origin_of_nonreciprocity}(c,d)]. This is required to conserve the phase relation between $m_v$ and $m_w$ (either advanced or retarded quadrature), which is imposed by the precessional motion. It follows immediately that, in one case, the in-plane component of $\mathbf{m}$ circulates around the extrema of $m_v$ [Fig.~\ref{Fig_Origin_of_nonreciprocity}(c)] so that a high degree of flux closure and little magnetic charges are achieved [Fig.~\ref{Fig_Origin_of_nonreciprocity}(a)]. In the other case, it does not [Fig.~\ref{Fig_Origin_of_nonreciprocity}(d)] and the magnetization strongly diverges on either side of the nodes of $m_v$, thus creating large amounts of magnetic charges there [Fig.~\ref{Fig_Origin_of_nonreciprocity}(b)].

As illustrated in Fig.~\ref{Fig_Origin_of_nonreciprocity}(a,b), channeled spin-waves correspond to a transverse vibration of the domain wall: the latter is periodically (in both time and space) displaced laterally, away from its equilibrium position. This oscillatory displacement gives rise to a change in the vertical component of magnetization at the equilibrium position of the wall center ($m_v$) and is accompanied by a change of the internal magnetization angle reflecting in the appearance of a magnetization component perpendicular to the wall ($m_w$). Thus, when spin-waves propagate, the Bloch wall essentially vibrates like a string. As emphasized in previous works\,\cite{MKT12,ZT18}, this string is however unusual in the sense that its wave equation violates the symmetries of time reversal and $u\!\mapsto\!-u$ mirror reflection and that it possesses gyrotropic properties, which result in transverse waves having different velocities depending on their direction of propagation.

Let us now analyze how the frequency non-reciprocity of the DWCSW mode depends on the details of the equilibrium configuration. i) On reflecting the system about the $uw$ midplane, the equilibrium magnetization $\mathbf{M}_{\text{eq}}$ is reversed at the wall center but is left unaffected in the domains. Moreover, the wave vector $\mathbf{k}\!=\!k\mathbf{u}$ is unchanged and, importantly, both in-plane components of $\mathbf{m}$, whose relative orientation determines the amount of dynamic magnetic charges created on spin-wave propagation [Fig.~\ref{Fig_Origin_of_nonreciprocity}(c,d)], reverse concomitantly. The frequency of the DWCSW mode must therefore be independent of the orientation of $\mathbf{M}_{\text{eq}}$ at the wall center\,\cite{Rem2}. ii) On rotating the system by $180^{\text{o}}$ about the $v$ axis, on the other hand, the direction of magnetization switches throughout the whole equilibrium configuration and $\mathbf{k}$ is reversed. Combining points i) and ii), we may conclude that the non-reciprocity of the DWCSW mode is solely determined by the orientation of magnetization in the stripe domains and that it is reversed when moving from a down/up ($\downarrow\uparrow$) configuration, as considered so far, to an up/down ($\uparrow\downarrow$) configuration: the dispersion relation of the DWCSW mode obeys $\omega_{\uparrow\downarrow}(k) =\omega_{\downarrow\uparrow}(-k)$ and where the channeled spin-waves have a larger (resp. smaller) wavelength in one case, they have a smaller (resp. larger) wavelength in the other case. Overall, we observe that the channeled spin-wave with smallest frequency obeys the following right hand thumb rule: it is the wave that travels in the direction pointed by the thumb as one wraps the right hand around the Bloch wall, with the extremal phalanges of the other four fingers and the palm oriented along the equilibrium direction of magnetization inside the stripe domains [Fig. 2(e)]. All this is consistent with the non-reciprocity being imprinted in the morphology of the non collinear equilibrium magnetic configuration and determined by the way reflectional symmetry about the $uv$ midplane is broken. We note that, to some extent, the non-reciprocity discussed here is similar to the magnetochiral spin-wave non-reciprocity predicted in ferromagnetic nanotubes with azimutal magnetic configurations\,\cite{OYSH16}.

\begin{figure}[h]
\includegraphics[width=8.5cm,trim=28 92 26 70,clip]{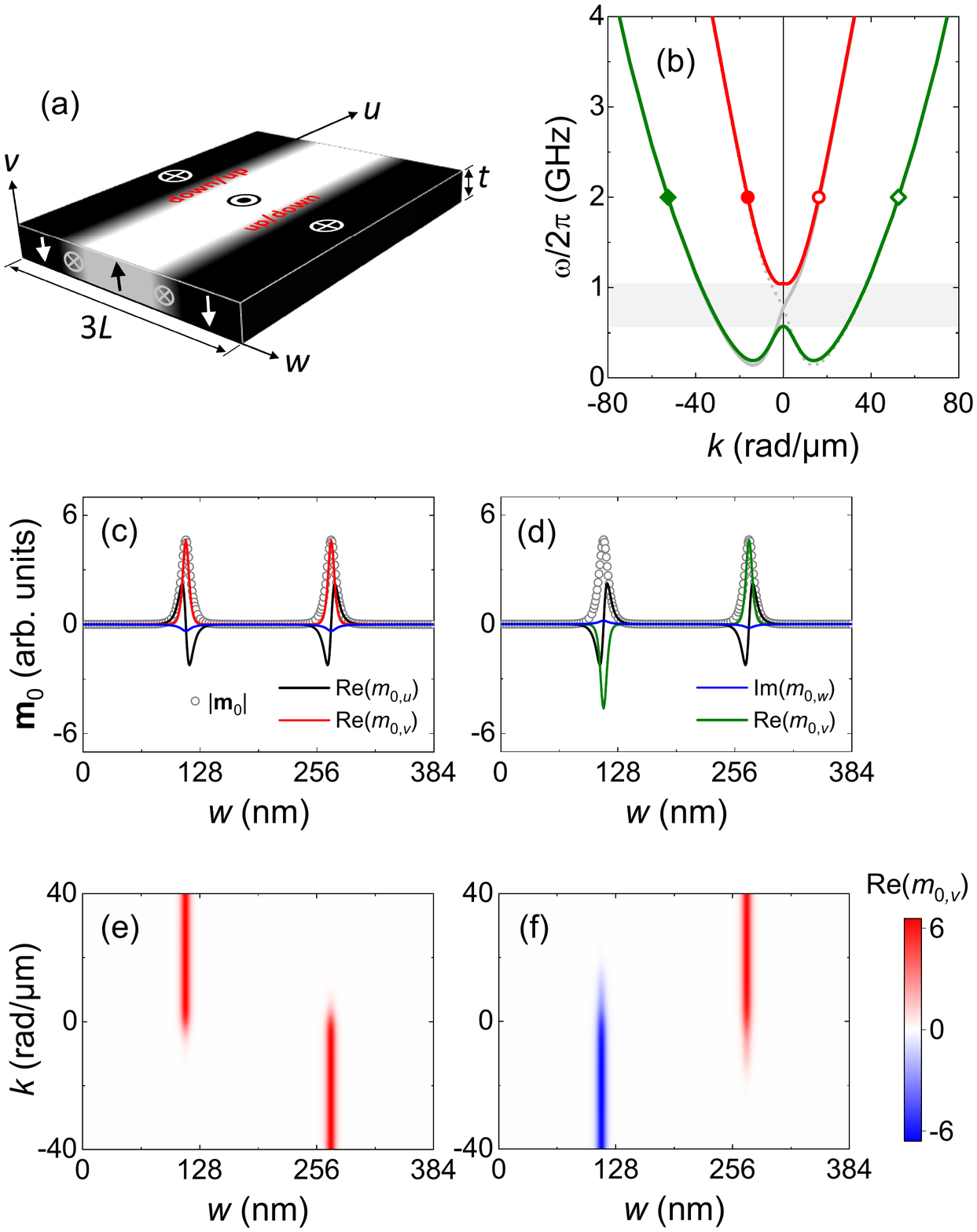}
\caption{(a) Schematic of the magnetic medium containing three stripe domains separated by two Bloch walls ($W\!=\!3L\!=\!384$~nm, $T\!=\!15$~nm). (b) Dispersion relations of the two collective DWCSW eigenmodes and (c,d) width profiles of the variable magnetization in those modes, for $k\!=0$. The mode with higher frequency (red lines) has an optic character while that with lower frequency (green lines) has an acoustic nature. For the purpose of comparison, the dispersion relations of DWCSW modes hosted by single down/up (solid grey line) and up/down (dashed grey line) walls in a $2L$-wide medium are also shown in panel (b). (e,f) Color maps of the amplitude of the vertical component of the variable magnetization, $\text{Re}(m_{0,v})$, in the optic-like (e) and acoustic-like (f) collective DWCSW eigenmodes as a function of the transverse coordinate $w$ and wave vector $k$.} \label{Fig_Two_Walls}
\end{figure}

We now turn our attention to the case of a magnetic medium with $W=3L=384$~nm containing three stripe domains and two Bloch walls, such that the magnetization in the central and outermost domains is pointing up and down, respectively [Fig.~\ref{Fig_Two_Walls}(a)]. This is of course the simplest possible wall array one can imagine. Here, the two domain walls are necessarily of opposite types: one is down/up and the other is up/down. Consequently, the situation is somewhat analogous to that of two inequivalent harmonic oscillators coupled together, as described classically\,\cite{N10}. Indeed, as seen before, the dispersion relations of the DWCSW modes bound to uncoupled walls of opposite types are symmetric to each other. Therefore, they intersect in $k\!=\!0$. On introducing dipolar coupling between the walls, as in our three stripe domain configuration, this degeneracy is lifted: frequency repulsion\,\cite{FB94} occurs and the dispersion relations of the two DWCSW modes no longer cross each other. Instead, an avoided crossing is formed around $k\!=\!0$, where the dipolar coupling is resonant [Fig.~\ref{Fig_Two_Walls}(b)]. Following how the profile of each of the two collective DWCSW modes evolves with $k$ reveals that the analogy with the case of two coupled harmonic oscillators can be taken even further. Near resonance (small $k$ values), the two modes are hybrid modes with non-zero amplitude at the locations of the two walls. One mode has an optic character since magnetization precession is in phase in the two walls [Fig.~\ref{Fig_Two_Walls}(c)] and wall displacements are therefore out-of-phase\,\cite{Rem3}, whereas the other one has an acoustic character [Fig.~\ref{Fig_Two_Walls}(d)]. Far from resonance (large $k$ values), that is, for dipolar decoupled walls, the normal modes retrieve a pure character: their amplitude is entirely localized within one wall and their frequency is close to that of a DWCSW mode bound to a single wall. Yet, when tuning $k$ from a large positive value to a large negative one through resonance (in a thought experiment), we observe that, for both modes, the amplitude is progressively transferred from one wall to the other [Fig.~\ref{Fig_Two_Walls}(e,f)]. This corresponds to a so-called adiabatic transition, whereby energy is transferred from one oscillator to the other\,\cite{N10}. As we will see below, this behavior has remarkable implications.

\begin{figure}[t]
\includegraphics[width=8.5cm,trim=101 120 47 90,clip]{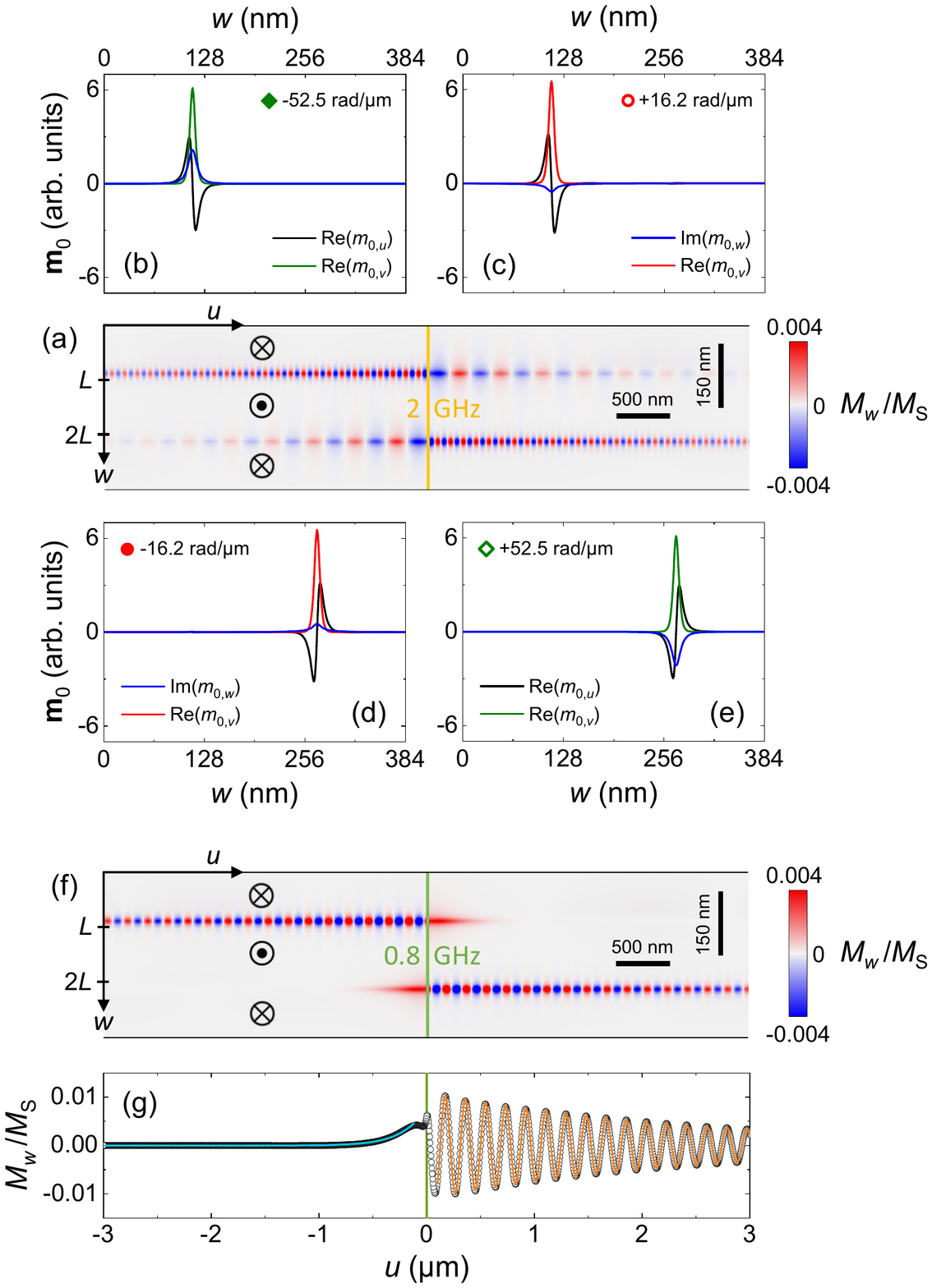}
\caption{(a) Map of the transverse component of the magnetization, $M_w$, as produced upon application of a localized (yellow line), vertically oriented, 5~mT alternating magnetic field with frequency $\omega/2\pi=2$~GHz to the three-stripe-domain medium sketched in Fig.~\ref{Fig_Two_Walls}(a). (b-e) Width profiles of the variable magnetization in the DWCSW eigenmodes involved in (a). The same color code is used as in Fig.~\ref{Fig_Two_Walls}(b-d). (f) Map of the transverse component of magnetization as generated upon application of a 5~mT alternating magnetic field with frequency $\omega/2\pi\!=0.8$~GHz (green line). (g) Line profile of $M_w$ at the equilibrium position of the up/down domain wall ($w\!\simeq\!2L$). Beyond a certain distance from the excitation source ($u\!=\!0$), the data (symbols) can be fitted (lines) to the expressions $M_w(u) = M_{0,w}\, \sin[k(u\!-\!u_0)]\, e^{-(u-u_0)/L_{\text{att}}}$ with $k=33.6$~rad/$\mu$m and $L_{\text{att}}=2.56~\mu$m, for $u\!>\!0$, and $M_w(u) = M_{0,w}\; e^{-(u-u_0)/L_{\text{att}}}$ with $L_{\text{att}}=0.21~\mu$m, for $u\!<\!0$.} \label{Fig_Two_Walls_2}
\end{figure}
\begin{figure}[h]
\includegraphics[width=8.5cm,trim=27 30 32 30,clip]{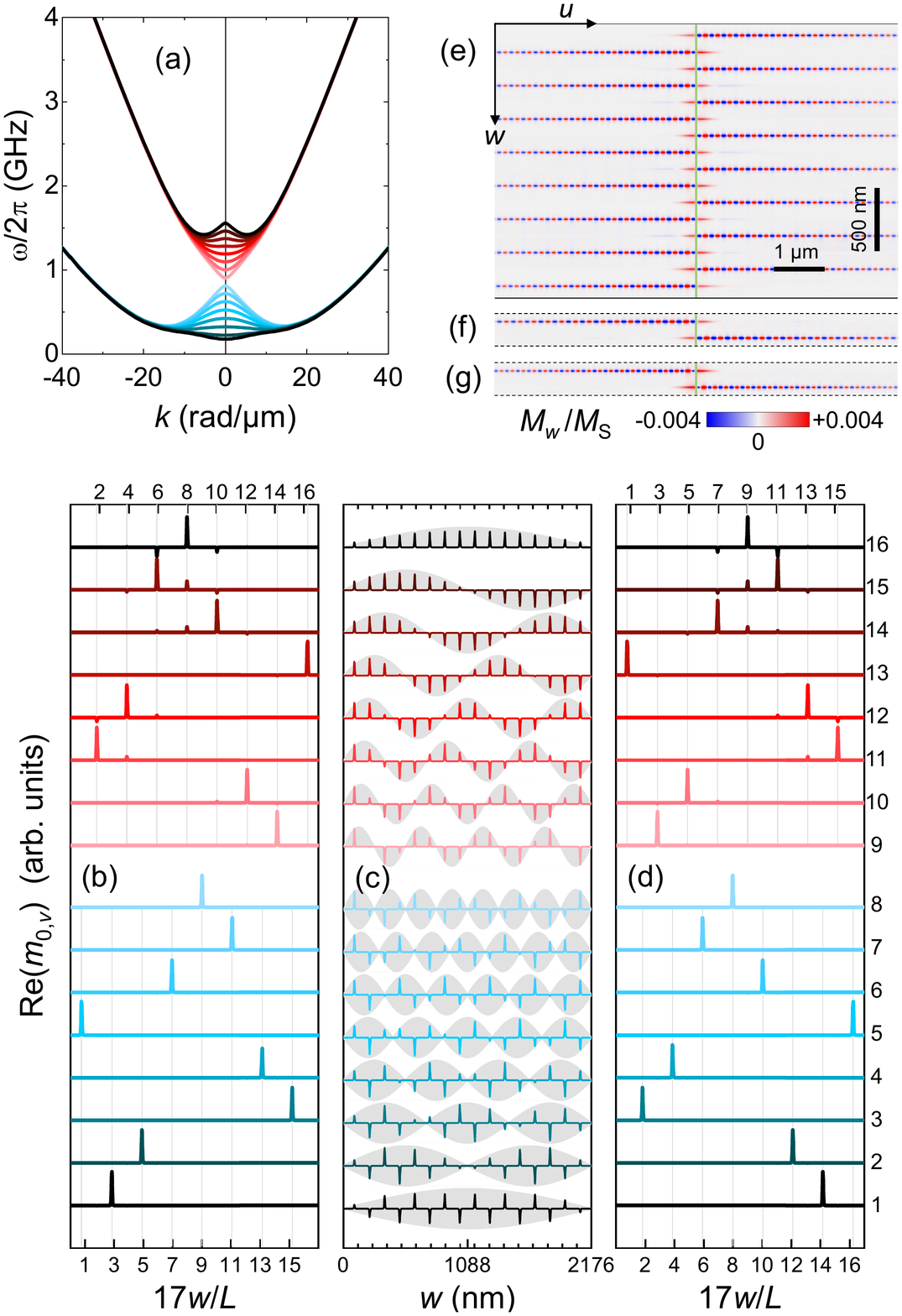}
\caption{(a) Dispersion relations of the DWCSW modes in a magnetic medium containing $N\!=\!16$ Bloch walls ($W\!=\!2176$~nm). The eight lowest (resp. highest) frequency modes, whose $\omega(k)$ curves are shown as blue lines (resp. red lines) of varying shade, have an acoustic (resp. optic) character. (b-d) Width profiles of the out-of-plane component of the dynamic magnetization in the sixteen DWCSW modes, for three different values of the wave vector: (b) $k\!=\!-30$~rad/$\mu$m, (c) $k\!=\!0$~rad/$\mu$m, and (d) $k\!=\!+30$~rad/$\mu$m. For clarity, the profiles are offset vertically according to the mode indices. (e-g) Maps of the transverse magnetization component ($M_w$) upon application of a localized (green line), vertically oriented, 5~mT alternating magnetic field to a magnetic medium with $N\!=\!16$: (e) Full domain wall array for $\omega/2\pi=0.9$~GHz and zoom in the region containing the two innermost domain walls for (f) $\omega/2\pi=0.7$~GHz and (g) $\omega/2\pi=1.1$~GHz.} \label{Fig_Sixteen_Walls}
\end{figure}

Expectedly, if the two DWCSW modes are excited simultaneously, using for instance a localized 2~GHz alternating magnetic field, spin-waves are produced, which propagate in both directions along the two walls [Fig.~\ref{Fig_Two_Walls_2}(a)]. What is not trivial and is revealed by our normal mode analysis is that, unlike in the case of a single wall, waves travelling in opposite directions inside a particular wall correspond here to different eigenmodes [Figs.~\ref{Fig_Two_Walls}(b) and \ref{Fig_Two_Walls_2}(b-e)]. This property makes it possible for Bloch walls to behave individually as spin-wave diodes or unidirectional magnonic waveguides. Indeed, if only the lowest frequency DWCSW mode is now excited using an alternating magnetic field with a frequency in the gap formed between the two DWCSW branches [see grey area in Fig.~\ref{Fig_Two_Walls}(b)] then propagating spin-waves are only launched in one direction, to the left ($u<0$) in the down/up wall and to the right ($u>0$) in the up/down wall [Fig.~\ref{Fig_Two_Walls_2}(f)]. In the forbidden direction of propagation, the responses of the domain walls to the excitation take the form of laterally confined evanescent waves, which decay extremely fast [Fig.~\ref{Fig_Two_Walls_2}(g)].

A pair of Bloch walls of opposite types, as considered above, constitutes a basic building block, whose behavior allows one to understand qualitatively that of much bigger arrays, with many such blocks. To illustrate this point, we finally consider a magnetic medium with $W\!=\!17L\!=\!2176$~nm containing sixteen Bloch walls. For such a system, there exists sixteen DWCSW modes (as many as walls), organized in a rich magnonic band structure [Fig.~\ref{Fig_Sixteen_Walls}(a)]. In particular, a hierarchy of almost equally spaced spin-wave branches is formed in the long wavelength limit. From the anatomy of the corresponding eigenmodes [Fig.~\ref{Fig_Sixteen_Walls}(c)], it is clear that the latter arises from the combined effect of i) dynamic dipolar interactions, which yield optic-like and acoustic-like modes, in equal numbers, and ii) lateral confinement related to the finite size of the magnetic medium, which introduces spin-wave quantization and lateral standing wave pattern in the mode profiles\,\cite{MJFD98,WKNL02}. With increasing $|k|$, dynamic dipolar coupling decreases, and so does the frequency splitting between all the branches of the same nature, located either above (optic-like branches) or below (acoustic-like branches) the avoided crossing. At large $k$, the amplitude of each mode eventually localizes in a unique wall of a particular type (or index parity), either down/up (odd index) or up/down (even index) depending both on the nature of the mode and on the sign of $k$ [Fig.~\ref{Fig_Sixteen_Walls}(b,d)]. The DWCSW modes then form two very distinct groups, with all the modes in a group being almost degenerate in frequency and each group corresponding to spin-wave channeling in one in every two walls. As in the case of a simple pair of walls, modal amplitude is transferred from a wall of type $A$ to a wall of type $B$ on reversing $\mathbf{k}$ and waves that travel in opposite directions inside a particular wall systematically belong to eigenmodes of different natures. If only acoustic-like modes are excited, individual Bloch walls behave once again as spin-wave diodes [Fig.~\ref{Fig_Sixteen_Walls}(e)]. This proves the generality of the phenomenon.

With a view to the possible practical use of this property in applications, it must be noted that the frequency gap between optic- and acoustic-like spin-wave branches becomes extremely small as the number of domain walls is increased (95~MHz for $N\!=\!16$). In practice, however, unidirectional spin-wave propagation is not necessarily restricted to this narrow frequency range. Indeed, the normal modes closest in frequency to the gap have profiles such that they hardly couple to laterally homogeneous excitations produced by magnetic field line sources, as those assumed for our numerical simulations, or coplanar waveguide inductive antenna, as most often employed in experiments\,\cite{BOF03,CDMA16}. As a result, the effectively useful frequency window can be much wider, reaching nearly half a GHz in the present case [Fig.~\ref{Fig_Sixteen_Walls}(f,g)]. Such should also be the case for a wall array with infinite or macroscopic size.

Logic architectures based on the manipulation of spin-wave amplitude would benefit from the existence of unidirectional spin-wave emitters\,\cite{XZ16,JKSL13}. This is the reason why the design of diode-like spin-wave devices has recently been addressed by the magnonics community\,\cite{LYWX15,BBGP17}. In the present work, we have demonstrated that, in addition to acting as self-cladding magnonic waveguides, conventional Bloch walls could behave just like this, provided one is capable of addressing individual walls in a dense array. Here, nonreciprocal propagation/emission of spin-waves does not require wave-vector selective excitation sources\,\cite{BBGP17} and does not rely on the interfacial Dzyaloshinskii-Moriya interaction, the effect of which is sufficiently strong in ultrathin films only. It occurs as a consequence of a collective effect and of conventional dynamic dipolar interactions within the symmetry broken spin textures of Bloch walls, which have the advantage of being hosted by magnetic media with thickness in the tens of nanometer range, maybe better suited for future applications. We note that the interest of such walls is further reinforced by the fact that strong dipolar-induced non-reciprocity effects, including unidirectional spin-wave propagation, survive when the wall structure departs from the perfect Bloch geometry and evolves towards a flux closure domain wall structure, through the formation of N\'eel caps, as usually observed in materials with larger thickness and/or smaller $Q$ factor\,\cite{HS98}.

\begin{acknowledgments}
The authors acknowledge fruitful discussions with Robert Stamps and financial support from the French National Research Agency (ANR) under Contract No.~ANR-16-CE24-0027 (SWANGATE).
\end{acknowledgments}


\begin{thebibliography}{0}
\bibitem{Rem1}
In such circumstances, the off-diagonal elements of the magnetostatic Green's function for plane spin-waves, which change sign when reversing $\mathbf{k}$, play a significant role. See Ref.~\onlinecite{K13}.
\bibitem{Rem2}
This shows that non-reciprocity does not depend on the chirality of the Bloch wall. See Ref.~\onlinecite{HGB16}.
\bibitem{Rem3}
The out-of-plane component of the dynamic magnetization is related to the lateral displacement of the wall $\delta w$ through  $m_v\!=\!-\frac{\partial M_{\text{eq},v}}{\partial w}\,\delta w$, where $\mathbf{M}_{\text{eq}}$ denotes the equilibrium magnetization vector. Adjacent walls in a stripe domain structure being necessarily of opposite types, they are characterized by opposite signs of $\frac{\partial M_{\text{eq},v}}{\partial w}$. It follows that in-phase (resp. out-of-phase) magnetization precession in neighboring walls corresponds to out-of-phase (resp. in-phase) displacements of those walls.
\end{thebibliography}
\end{document}